\documentclass[twocolumn,secnumarabic,amssymb,nobibnotes,aps,prd,showpacs]{revtex4-1}

\usepackage[pdftex]{graphicx}

\begin{document}

\title{Infrared Dielectric Resonator Metamaterial}%

\author{James C. Ginn}
\altaffiliation{Present Address: Plasmonics Inc, Orlando, FL 32826, USA.}
\email[Electronic Mail: ]{james.ginn@plasmonics-inc.com}
\author{Igal Brener}
\affiliation{Sandia National Laboratory, Albuquerque, NM 87185, USA.}
\affiliation{Center for Integrated Nanotechnologies, Sandia National Laboratory, Albuquerque, NM 87185, USA.}
\author{David W. Peters}
\author{Joel R. Wendt}
\author{Jeffrey O. Stevens}
\author{Paul F. Hines}
\author{Lorena I. Basilio}
\author{Larry K. Warne}
\author{Jon F. Ihlefeld}
\author{Paul G. Clem}
\author{Michael B. Sinclair}
\affiliation{Sandia National Laboratory, Albuquerque, NM 87185, USA.}
\date{\today}%

\begin{abstract}
We demonstrate, for the first time, an all-dielectric metamaterial resonator in the mid-wave infrared based on high-index tellurium cubic inclusions. Dielectric resonators are desirable compared to conventional metallo-dielectric metamaterials at optical frequencies as they are largely angular invariant, free of ohmic loss, and easily integrated into three-dimensional volumes. With these low-loss, isotropic elements, disruptive optical metamaterial designs, such as wide-angle lenses and cloaks, can be more easily realized.
\end{abstract}

\maketitle

The unique properties of metamaterials have yielded many exciting electromagnetic phenomena including sub-diffraction-limited imaging \cite{Pendry1}, cloaking \cite{Schurig1}, and perfect absorption \cite{Landy1}.  In spite of the rapid advances in this field, passive metamaterials at optical frequencies have often proven impractical due to significant conductor loss from the metallic resonators comprising these volumes \cite{Xiao1}. Additionally, the inherent geometrical asymmetry of these resonators further restricts metamaterial behaviour to a small range of incident angles even when assembled into three-dimensional structures \cite{Burckel1}.  Three-dimensional dielectric resonators, unlike their metallic counterpart, have significantly less material loss, support resonant modes that are invariant to the excitation angle, and can be easily integrated into thick volumes.  In this letter, we describe the development of a dielectric resonator based metamaterial in the infrared with spectral regions of negative magnetic and electric effective properties. Additional insight is also provided in addressing material limitations imposed on dielectric metamaterials at optical frequencies.

\begin{figure}
\includegraphics[scale=1]{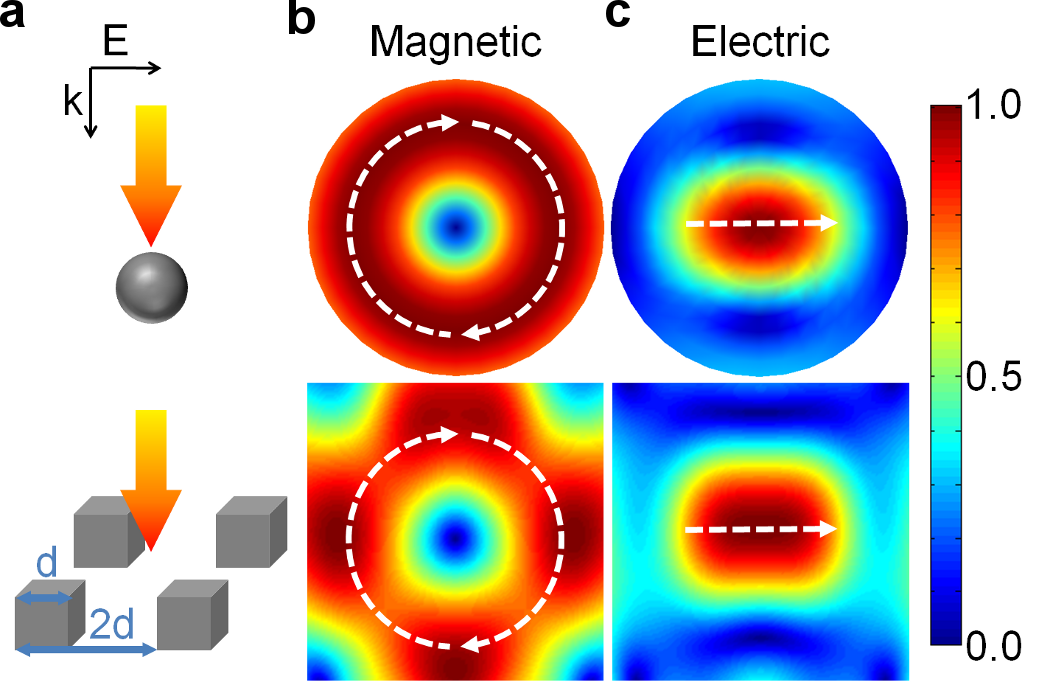}
\caption{(color online). (a) Excitation configuration of an isolated sphere (top row) and a 1:1 periodic cube array (bottom row). (b) normalized electric field distribution for the lowest-order magnetic and (c) normalized electric field distribution for the lowest-order electric mode in the sphere and cubic resonators. Dotted white lines indicate field direction.}
\label{fig:1}
\end{figure}

\begin{figure*}
\includegraphics[scale=1]{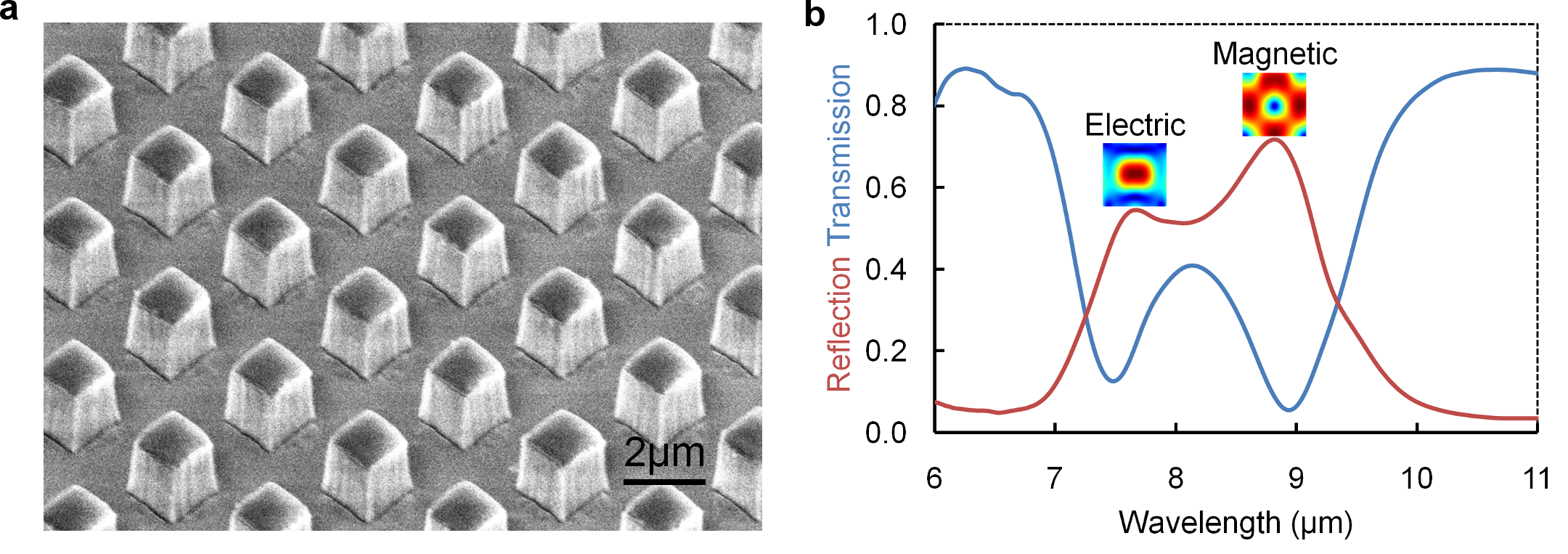}
\caption{(color online). (a) Scanning electron micrograph of fabricated CDR. (b) Measured reflection and transmission coefficients for CDR. Field patterns from Fig. \ref{fig:1} are shown above each corresponding resonance.}
\label{fig:2}
\end{figure*}

Lewin \cite{Lewin1} theoretically demonstrated that an array of sub-wavelength dielectric resonators can exhibit spectral regions of Lorentzian-like effective permittivity and permeability.  Following Mie theory \cite{Bohren1}, the effective electric and magnetic polarizabilities of a densely packed array of sub-wavelength spheres can be altered by changing the dimensions, composition, and packing fraction of the inclusions.  Similar behavior can be achieved using cubic dielectric resonators (CDRs) \cite{Zhao1} which, unlike spheres, are compatible with existing nano-scale lithography techniques and can be integrated into multi-layer composites through repeated steps of thin-film deposition, etching, and planarization.  Since there are no analytical expressions for predicting the behavior of a cubic scatterer \cite{Napper1} and approximations are problematic due to the high index dispersion of materials in the infrared, numerical computational electromagnetic approaches must be used for analysis.  In Fig. \ref{fig:1} the analytically determined on-resonance field distribution of an isolated sphere is compared to that of a cubic resonator array calculated using the commercially available rigorous coupled wave analysis (RCWA)\cite{Moharam1} package, GDCALC.  Like the spherical resonator, the lowest-order mode of a CDR is a magnetic dipole (TE$_{011}$) and the second-lowest mode is an electric dipole (TM$_{011}$).  However, in contrast to traditional metallic resonators the electromagnetic responses of single spherical inclusions are isotropic, and cubic resonators represent only a minor perturbation from this spherical symmetry.  Also, whereas significant damping due to ohmic loss is unavoidable for metallic resonators in the infrared \cite{Ginn1}, the damping of dielectric resonators can be quite low provided the resonator material lacks active carriers or phonon modes in the band of interest.  The scattering cross-sections exhibited by such undamped systems are known to be equal to the fundamental limits imposed by Mie theory \cite{Schuller1}.

Thus, identification of candidate low-loss resonator materials is critical in designing a practical infrared CDR metamaterial. In addition, the CDR material must possess a large index of refraction to ensure that the dimensions of the resonator and array spacing are sufficiently small (non-diffracting) compared to the operating wavelength. Only two classes of dielectrics exhibit positive indices of refraction greater than three in the infrared: highly crystalline polaritonic materials and narrow band-gap ($<$ 1.5 eV) materials. Metamaterial structures using polaritonic materials have previously been investigated \cite{Schuller2}, but these materials are less desirable due to their high loss at the phonon resonance and limited spectral flexibility.  In contrast, narrow band-gap materials exhibit large indices over wide spectral bands, and only experience significant loss near the band-gap at shorter wavelengths and on the tail of the free-carrier absorption at longer wavelengths.  Candidate narrow band-gap materials for infrared CDR designs include silicon \cite{Vynck1}, germanium, tellurium, and IV-VI compounds containing lead (such as lead telluride).

Initial investigations of a magnetic active CDR array in the infrared was carried out using germanium cubes on a low-index polymer thin-film \cite{Ginn2}. Tellurim (Te) was subsequently selected as a better resonator material due to its larger index of refraction and low loss at infrared wavelengths \cite{Fink1}.  Because of its trigonal crystal lattice \cite{Weidmann1}, Te is naturally anisotropic in single crystal form with an extraordinary index of refraction of 6.2 at 10 $\mu$m \cite{Palik1}. For CDR applications, a polycrystalline morphology is preferable which yields a crystal-averaged index of refraction of 5.3 at 10 $\mu$m, with an extinction coefficient of less than 10$^{-4}$ \cite{Palik1}. Barium fluoride (BaF$_{2}$) was selected as the optimal substrate due to its low refractive index (n $\sim$ 1.4) and low loss at 10 $\mu$m.

Through simulation, a 1.7 $\mu$m CDR with a 3.4 $\mu$m unit-cell spacing (1:1 duty-cycle) was chosen to center the reflection peak of the magnetic resonance at 10 $\mu$m.  Prior to patterning, a 1.7 $\mu$m thick film of Te was deposited on the surface of a 25 mm diameter BaF$_{2}$ optical flat via electron-beam evaporation.  Both x-ray diffraction and variable angle spectral ellipsometric analysis verified that the film was predominately polycrystalline, and ellipsometry analysis yielded a fitted complex index of refraction of n = 5.02 + 0.04j.  The Te film was patterned using electron beam lithography and etched using a reactive ion etching (RIE) process.  A scanning electron micrograph of the etched pattern is shown in Fig. \ref{fig:2}a.  The etching process resulted in excellent uniformity over a 1 cm$^{2}$ area, with only a slight over-etching of the pattern.  The final CDR element was 1.7 $\mu$m tall with a 1.53 x 1.53 $\mu$m base and a 10 degree sidewall slope.  The overall process required significantly less steps than existing three-dimensional metallo-dielectric lithography \cite{Burckel1} and the feasibility of planarizing the patterned region for multi-layer fabrication was verified by successfully spin-coating a thin-film of polynorbornene on the surface of the CDR array \cite{Rasberry1}. Although the CDRs have an isotropic resonance mode, multi-layer fabrication is necessary for an angular-independent array response.

Following fabrication, the patterned wafer was characterized using a hemispherical directional reflectometer (HDR) at an angle of incidence of seven degrees.  The measured collimated transmission and specular reflection of the array are plotted in Fig. \ref{fig:2}b.  As expected from the fabricated topology and measured index of the Te film, the reflection peak of the magnetic resonance occurred at 9 $\mu$m and the reflection peak of the electric resonance occurred at 7.5 $\mu$m.  The resonances are well defined and occur at wavelengths above the diffraction cut-off limit.  In the spectral region between the two resonances, loss (1 – reflection – transmission) drops to less than 8\%.  We note that no artificial correction factors were used to account for reflection and absorption losses due to the substrate.

\begin{figure}
\includegraphics[scale=1]{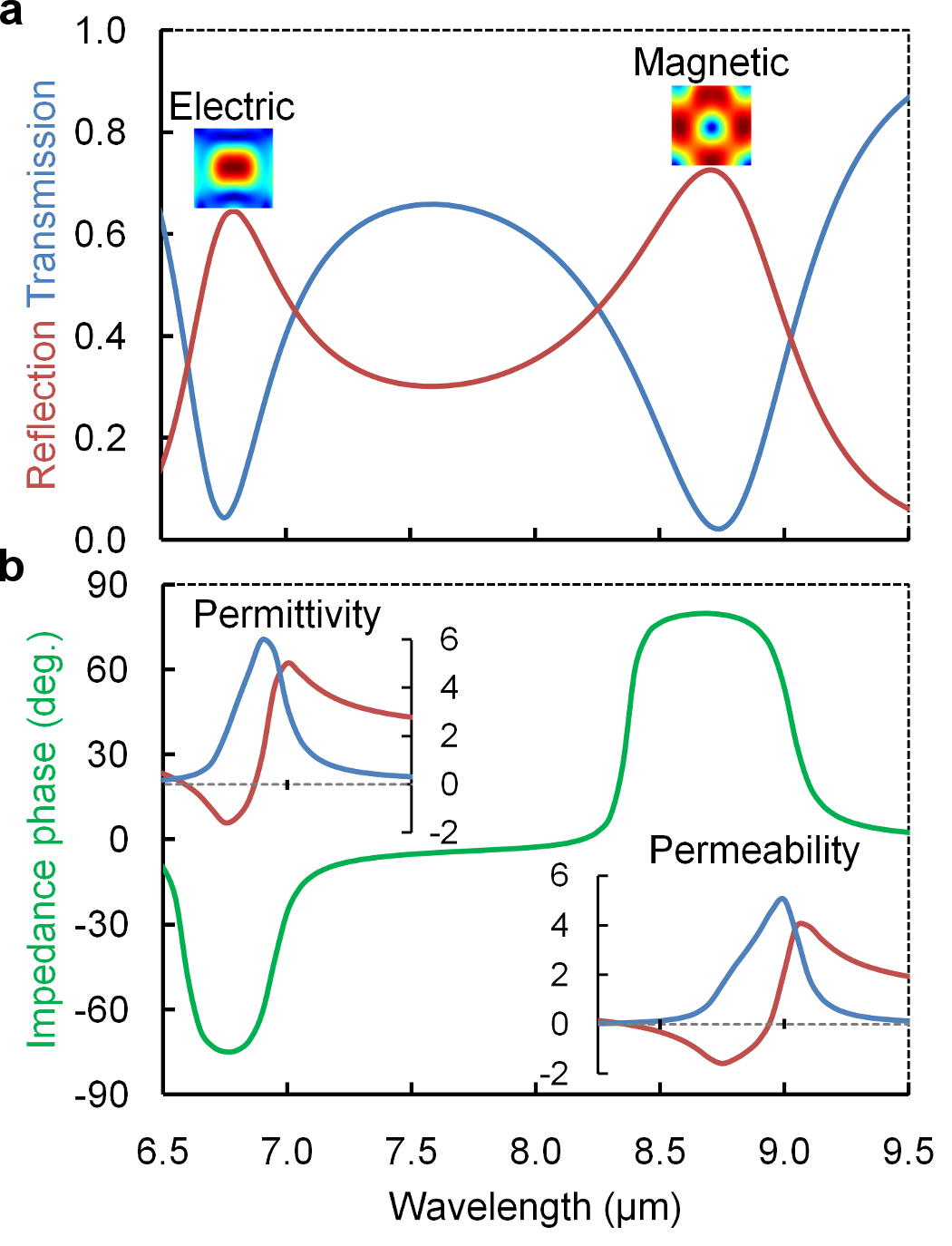}
\caption{(color online). (a) Plot of simulated reflection and transmission for CDR. (b) Plot of calculated impedance phase, permittivity, and permeability for the simulated CDR array. Real values are in red and imaginary values are in blue.}
\label{fig:3}
\end{figure}

The optical response of the fabricated cube array was simulated (Fig. \ref{fig:3}). The simulated and measured coefficients show overall good agreement (Fig. \ref{fig:3}a), with differences primarily arising from the asymmetry and non-uniformity of the as-fabricated cube.  The calculated surface impedance (Fig. \ref{fig:3}b) indicates two distinct resonances with inverted phase delays.  Positive phase delay occurs when the electric field leads the magnetic field in phase (permeability less than zero) and negative phase delay occurs when the electric field follows the magnetic field (permittivity less than zero).  Using a standard retrieval algorithm \cite{Smith1} for the planar array, the effective permittivity and permeability were calculated for these two regions (inset 3b).  In both cases, the extracted parameters reach values of less than -1, and the loss tangent of the permeability falls to 0.48 when the real part of permeability is equal to -1.  We note that the extinction coefficient of the as deposited film is more than two orders of magnitude larger than the literature value for Te.  Thus, we anticipate that significantly lower-loss metamaterials will be achievable as the Te loss is minimized.

\begin{figure}
\includegraphics[scale=1]{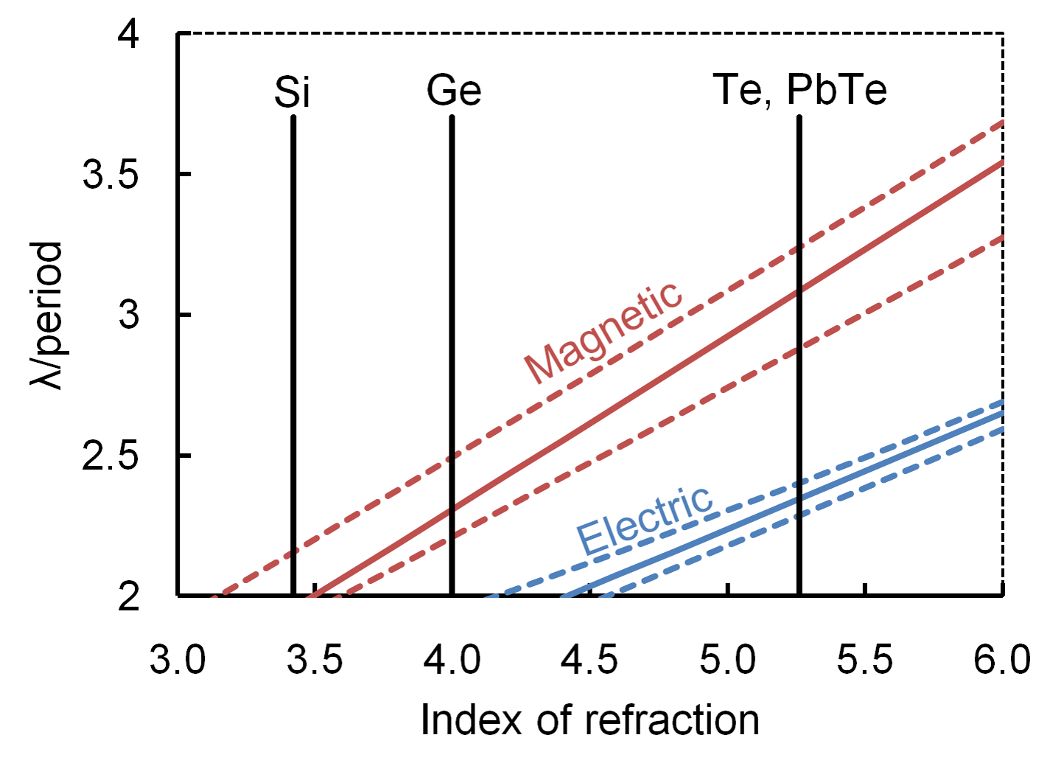}
\caption{(color online). Design metric for 1:1 CDR metamaterials. Solid lines correspond to the lowest-order magnetic (red) and electric (blue) resonances. The top dotted line defines where peak positive permeability/permittivity occurs and the bottom line defines where permeability/permittivity is equal to zero. The indices of several materials at 10 $\mu$m are also labeled.}
\label{fig:4}
\end{figure}

To further investigate the features of infrared CDRs, a series of RCWA simulations was run for a 1:1 duty-cycle array while only varying the refractive index of the cubes.  The solid lines in Fig. \ref{fig:4} denote the normalized wavelength at the points of peak reflectivity (due to the two primary resonances), versus index of refraction (restricted to a range of values that is realistic for known optical materials).  These lines also directly correspond to the point where the CDR array switches from supporting a propagating mode (positive effective permeability/permittivity) to a plasma mode (negative effective permeability/permittivity). The effective permeability and permittivity were also calculated using the retrieval algorithm \cite{Smith1} assuming a fixed extinction coefficient of 0.001 for algorithm stability.  From these calculated values, the points of peak effective positive permeability/permittivity and the zero crossing of the effective permeability/permittivity curves were determined and plotted as dotted lines on Fig. \ref{fig:4}.

Several critical performance metrics for optical CDRs can be gleaned from Fig. \ref{fig:4} and related analysis.  As expected, the resonant wavelength of a CDR will scale linearly with the dimensions and index of refraction of the cube. This behavior explicitly limits the practicality of negative permitivity metamaterials in the mid-wave infrared based around traditional high-index materials such as silicon and germanium. This limitation is also more severe in the optical regime when compared to the rf and THz portions of the spectrum where materials with indices surpassing thirty exist \cite{Zhao2}. These arrays also exhibit appreciable spatial dispersion \cite{Liu1} that decreases asymptotically with increasing normalized wavelength (the array period is decreasing relative to the resonant wavelength).  Spatial dispersion dominates when the effective index of the CDR array exceeds its normalized wavelength and field homogenization breaks down.  In this regime, a photonic crystal band-gap mode is excited and retrieved optical properties no longer hold physical meaning.  Consequently, the peak effective positive permeability/permittivity curve defines the largest index of refraction supported by the array at normal incidence and the start of the band-gap region.  This regime persists until the resonance line is crossed and the behavior of the CDR array becomes dominated by the plasma mode for which spatial dispersion can largely be ignored.  More exotic behaviors, such as doubly negative parameters, can be realized by mixing two dielectric resonators with different indices of refraction or different dimensions in a multi-layer composite. 

\begin{figure}
\includegraphics[scale=1]{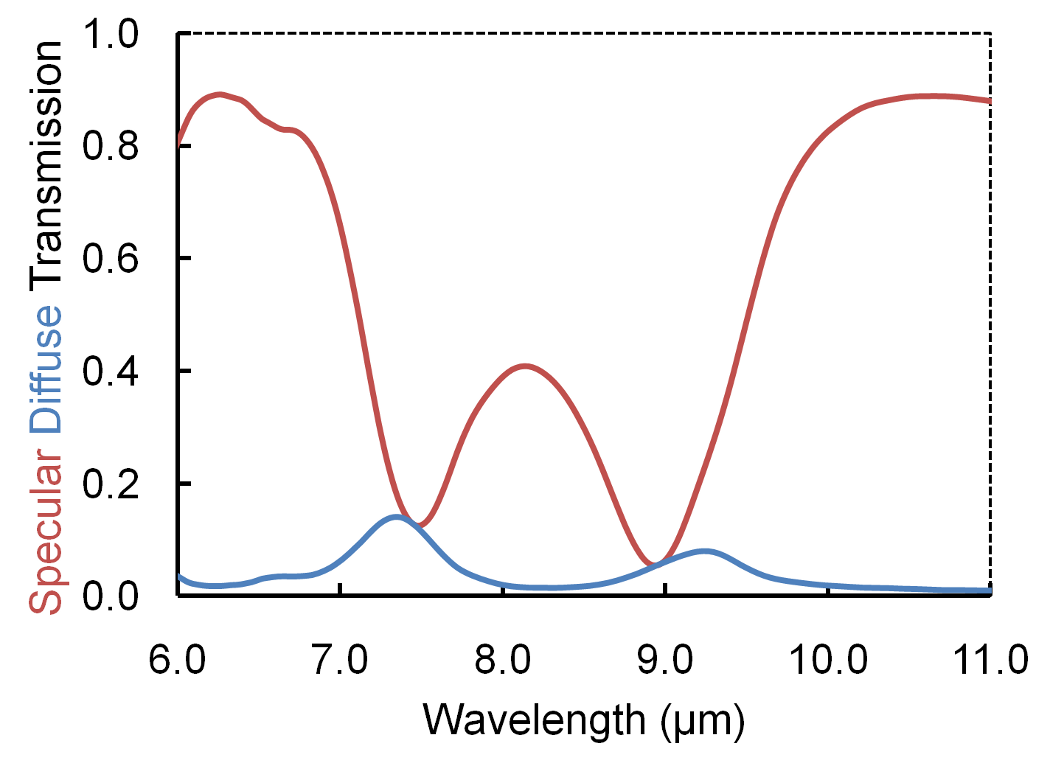}
\caption{(color online). Specular (red line) and diffuse (blue line) transmission for measured CDR.}
\label{fig:5}
\end{figure}

One of the advantages of characterizing the fabricated metamaterial with an HDR is that it allowed for experimental validation of the onset of spatial dispersion in the fabricated CDR. Fig. \ref{fig:5} shows a comparison of the measured specular and diffuse (diffracted) transmission. From the figure, the region where the device is dominated by the photonic crystal band-gap mode corresponds directly to the appearance of appreciable coupling of incident light into diffracted orders. This behavior also manifests near in the center of the metamaterial's resonance and is stronger for the electric mode, as expected from theory. Furthermore, this confirms the effective loss of the CDR is much larger than loss associated with material absorption.

In this letter, we have described the design, fabrication, and characterization of a dielectric cubic resonator metamaterial with electric and magnetic activity in the mid-infrared. Through theory and simulation, a generalized design approach for metamaterial surfaces comprising of cubic resonators at optical frequencies was developed. This work represents a first step toward the development of passive low-loss, multi-layer, isotropic metamaterial devices in the infrared.

This research was supported by the Laboratory Directed Research and Development program at Sandia National Laboratories.  This work was performed, in part, at the Center for Integrated Nanotechnologies, a U.S. Department of Energy, Office of Basic Energy Sciences user facility.  Sandia is a multi-program laboratory operated by Sandia Corporation, a Lockheed Martin Company, for the U.S. Department of Energy under contract DE-AC04-94AL85000.  

\bibliography{refs}

\end{document}